\documentclass[%
preprint,
 amsmath,amssymb,
 aps,
]{revtex4-2}

\usepackage{graphicx}
\usepackage{dcolumn}
\usepackage{bm}

\begin{document}

\newcommand\liho{LiHoF$_{4}$}
\newcommand\lihox{LiHo$_{x}$Y$_{1-x}$F$_{4}$}
\newcommand\lihofour{LiHo$_{0.045}$Y$_{0.955}$F$_{4}$}

\newcommand\ftwodf{f_{2 \Delta f}}
\newcommand\fpump{f_{\text{pump}}}
\newcommand\fprobe{f_{\text{probe}}}
\newcommand\hpump{h_{\text{pump}}}
\newcommand\hprobe{h_{\text{probe}}}
\newcommand\omegatwodf{\omega_{2 \Delta f}}
\newcommand\omegapump{\omega_{\text{pump}}}
\newcommand\omegaprobe{\omega_{\text{probe}}}
\newcommand\omegacars{\omega_{\text{CARS}}}
\newcommand\omegacsrs{\omega_{\text{CSRS}}}

\newcommand\chione{\chi^{(1)}}
\newcommand\chitwo{\chi^{(2)}}
\newcommand\chithree{\chi^{(3)}}
\newcommand\chifive{\chi^{(5)}}

\title{Four-Wave Mixing at p$e$V Energy Scales}

\author{C. Simon}
 \altaffiliation{Division of Physics, Mathematics, and Astronomy, California Institute of Technology, Pasadena, CA 91125}

\author{D. M. Silevitch}
 \altaffiliation{Division of Physics, Mathematics, and Astronomy, California Institute of Technology, Pasadena, CA 91125}

\author{T. F. Rosenbaum}
 \altaffiliation{Division of Physics, Mathematics, and Astronomy, California Institute of Technology, Pasadena, CA 91125}

\date{\today}

\begin{abstract}
We measure the non-linear magnetic susceptibility $\chithree$ of the disordered quantum Ising magnet, \lihofour, and demonstrate four-wave mixing due to coherent (anti-)Stokes Raman scattering at $\sim$ 100 Hz (p$e$V) energy scales. The temperature dependence of $\chithree$ approximately follows a $\frac{1}{T}$ form, with a high-$T$ cutoff that can be linked to dissipation in the coherent spin clusters. $\chithree$ also decreases monotonically with a transverse field, approaching a constant offset above a few kOe, suggesting the presence of both coherent, and spontaneous Raman scattering.
\end{abstract}

\maketitle


\section{Introduction}

Non-linear optical phenomena provide powerful insights into the internal structure of atoms, molecules, and solids. When optically-active materials are driven sufficiently strongly, well into the non-linear regime, exotic phenomena such as four-wave mixing \cite{oudar1980phys.rev.a, du2007phys.rev.lett.}, electromagnetically-induced transparency \cite{boller1991phys.rev.lett., liu2017nanophotonics}, and coherent population trapping \cite{agapev1993phys.-usp., xu2008naturephys} emerge. Of particular recent interest has been the stimulation of new, non-equilibrium states in Floquet-driven quantum materials via optical pumping \cite{zhang2017nature,choi2017nature,xu2020phys.rev.lett.,rudner2020natrevphys}. 

Intense laser excitation at optical energy scales of order $\sim$ $e$V \cite{2017} is the usual approach employed to access the non-linear response. We demonstrate here a magnetic analogue to non-linear optics via pump-probe experiments that use no more than a coil wrapped around a magnetic solid. The system in question is the highly-disordered quantum magnet, \lihofour, a physical realization of the Ising model in transverse field. In the dilute limit, where non-magnetic yttrium largely replaces the magnetic holmium dipoles without disturbing the structural integrity of the single-crystal material, we show that dipole clusters consisting of hundreds of spins \cite{ghosh2002science} exhibit four-wave mixing at p$e$V energy scales.

\begin{figure}
    \includegraphics[width=0.4\textwidth]{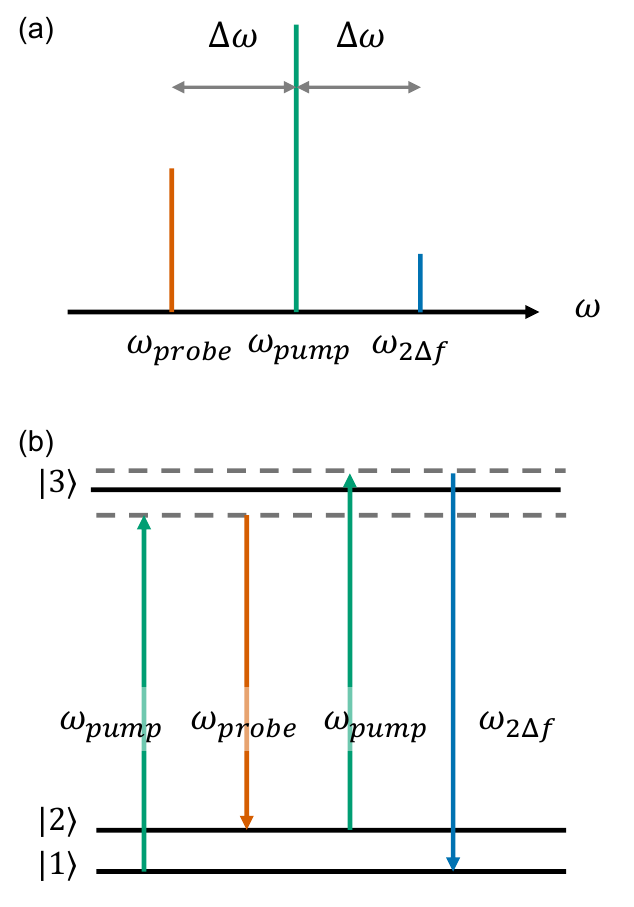}
    \caption{\label{fig:fig_1} Illustration of four-wave mixing process. (a) Third-order non-linear susceptibility, $\chithree$, generates a new spectral component at frequency $\omegatwodf = \omegaprobe + 2 \Delta \omega$ with $\Delta \omega \equiv \omegapump - \omegaprobe$. This response at $\omegatwodf$ cannot be generated by the linear susceptibility, $\chione$, and is inherently a non-linear effect. (b) For three-level systems on-resonance with the pump and probe frequencies, photons at $\omegatwodf$ are generated due to coherent (anti-)Stokes Raman scattering. In \lihofour, these coherent (anti-)Stokes Raman scattering processes occur with energies at the p$e$V energy scale.}
\end{figure}

The third-order non-linear susceptibility, $\chithree$, is known to cause four-wave mixing, generating new spectral components at frequencies other than the incident frequencies. We illustrate this process in Fig. \ref{fig:fig_1} (a), where the sample response exhibits not only spectral components at the incident photon (in our case, localized magnetic excitation) energies, $\omegaprobe$ and $\omegapump$, but also at a new energy $\omegatwodf$. Spectral components with frequency $\omegatwodf \neq \omegaprobe,\omegapump$ cannot be generated by the linear susceptibility $\chione$ – they are clear indicators of non-linearities in the sample response.

Previous pump-probe spectroscopic measurements on \lihofour $ $ demonstrated quantum interference between excitation pathways due to spin clusters with effective low-energy Lambda schemes that are nearly resonant with the pump and probe frequencies \cite{buchhold2020phys.rev.b}. Hence, the four-wave mixing processes in \lihofour $ $ can be attributed to coherent (anti-)Stokes Raman scattering processes, which we illustrate in Fig. \ref{fig:fig_1} (b).

\section{Experimental Methods}

We measured the magnetic susceptibility using a pump-probe technique. The 5 mm x 5 mm x 10 mm single crystal of \lihofour $ $ was pumped via an AC magnetic field at frequency $\fpump \sim$ 100 Hz, with amplitude $\hpump \sim$ 300 mOe, and scanned with a smaller probe field ($\hprobe \sim$ 30 mOe) at range of frequencies $\fprobe = \fpump \pm \Delta f$ with detuning $\Delta f \sim$ 1-10 mHz. We monitored the magnetic response of the sample via an inductive pickup coil, whose induced voltage was passed into a lock-in amplifier, which mixed down the signal by $\fprobe$. We were able to record not only the longitudinal response parallel to the AC pump and probe fields, but also the signal transverse to the Ising axis, by means of a vector susceptometer. We confirmed that the non-linear response arose from the sample itself by the null signal in an empty reference coil configured in the identical instrumentation chain.

The heat-sinking of the sample is crucially important. To experimentally access the required mK temperatures, the sample, in the susceptometer, was mounted on the cold finger of a helium dilution refrigerator. While the susceptometer itself was machined out of oxygen-free high purity copper (OFHC), the coil forms were machined out of Hysol epoxy, which is a poor thermal conductor. The thermal boundary conditions then can be tuned from a weakly-coupled regime to a strongly-coupled regime if the sample is additionally heat-sunk to a sapphire plate. Depending on the thermal boundary conditions, \lihofour $ $ will exhibit either a glassy response (if strongly coupled to the environment), or an ``anti-glass'' response (if weakly coupled to the environment) \cite{schmidt2014proc.natl.acad.sci.}. Since we are probing excitations at $\sim$ 100 Hz energy scales ($\approx$ 30 nK), even though the sample is cooled into the mK regime, we are still effectively in the infinite temperature regime. Consequently, the dissipation due to inelastic phonon scattering at these low energy scales is highly dependent on the phonon linewidth, $\Gamma$, which is determined by how well or poorly the sample is heat-sunk \cite{ buchhold2020phys.rev.b}.

The Fano resonances measured in \cite{schmidt2014proc.natl.acad.sci.,silevitch2019natcommun} were only visible in the regime of weak heat-sinking (thermal contact primarily mediated through the epoxy). Increased dissipation broadens the linewidths of the cluster excitations to the point that it destroys the Fano response. The measurements in this work follow the same weak heat-sinking conditions as in \cite{silevitch2019natcommun}.

\begin{figure}
    \includegraphics[width=0.4\textwidth]{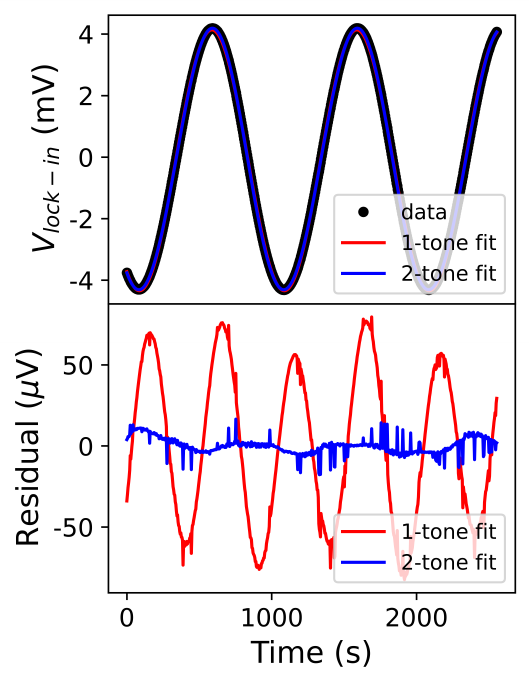}
    \caption{\label{fig:fig_2} Detection of non-linear susceptibility $\chithree$.The data is shown fit with both a one-tone sine at $f = \Delta f$ and a two-tone sine with extra component $f = 2 \Delta f$. While the magnitude of the spectral component at $2 \Delta f$ is two orders of magnitude smaller than that at $\Delta f$, it can be seen clearly in the residuals of a single-tone fit.}
\end{figure}

Since the detuning is small enough to survive the lock-in low-pass filtering, the output must be filtered a second time, digitally, in order to separate the various spectral components. To extract each spectral component, the data was fit to a multi-tone sine wave. First, the raw data were fit to a two-tone sine fit using linear least-squares regression. We plot a sample fit in Fig. \ref{fig:fig_2}. As the residual from a single-tone sine fit in Fig. \ref{fig:fig_2} shows, there is a clear signal in the lock-in output at $2 \Delta f$, albeit two orders of magnitude smaller than the signal at $\Delta f$.

After the lock-in amplifier mixes down the signals down by $\fprobe$, the sample response at the probe frequency will appear as a DC component in the output, while the sample response at $\fpump$ appears as a signal at $\pm \Delta f$. Previous work employing this technique \cite{silevitch2019natcommun,schmidt2014proc.natl.acad.sci.} focused exclusively on the signal at $\fprobe$, discarding the other spectral components. The signal at $\fprobe$ displays asymmetric Fano resonances \cite{silevitch2019natcommun}, a clear signature of quantum interference. 

In general, the magnetic susceptibility, $\chi$, of any material, which relates its magnetization, $M$, in response to an external magnetic field, $H$, can be Taylor-expanded in powers of the external field: $M_\alpha = \chione_{\alpha\beta} H_\beta + \chitwo_{\alpha\beta\mu} H_\beta H_\mu + \chithree_{\alpha\beta\mu\nu} H_\beta H_\mu H_\nu + \ldots$. However, the inversion symmetry of \lihox\ precludes all even terms in this expansion, leaving us with only odd powers: $\chi = \chione + \chithree + \chifive + \ldots$. Since $\chione$ is linear in the external field, it cannot generate a response at frequencies other than the spectral components present in the external AC field. However, non-linear terms, such as $\chithree$, can generate not only harmonics of each spectral component, but also create new frequencies at sums/differences of spectral components of the external field.

Given that the sample is only driven at the two frequencies, $\fprobe$ and $\fpump$, which show up as signals at $f=0$ or $\Delta f$ in the lock-in output, any other spectral component than 0 or $\Delta f$ must be a result of sample non-linearities. We focus on the spectral component of the sample response at $\ftwodf \equiv \fprobe + 2 \Delta f$. This new frequency can be generated by four-wave mixing, in which the third-order susceptibility, $\chithree$, takes in two incident photons and inelastically scatters them into two photons with different energies. More specifically, this process can take in two photons of energy $\omegapump$ and scatter them into one photon of energy $\omegaprobe$ and another photon of energy $\omegatwodf$. While in principle higher order terms, such as $\chifive$, can also generate photons at $\omegatwodf$, so any signal at $\ftwodf$ is in theory due to a sum of all of these higher order processes, the strength of these higher order terms falls off rapidly. We restrict the discussion of our data to the leading order non-zero term that can generate a spectral response at $\omegatwodf$, $\chithree$.

It is possible to probe the non-linear susceptibility by considerably increasing the amplitude of the drive field, rather than looking at new frequency generation. However, at the excitation levels required, varying the pump amplitude also varies the effective sample temperature $T$ at dilution refrigerator temperatures due to eddy-current heating generated in the copper susceptometer. Such an approach convolves the sample non-linear behavior of interest with the effects of heating, which cannot be separated easily. Furthermore, while previous measurements \cite{silevitch2007phys.rev.lett., silevitch2019natcommun} probe the non-linear dynamics of \lihofour, the signal at the probe frequency is a convolution of both linear and non-linear processes. For these reasons, we rely on the four-wave mixing technique to characterize directly the non-linear response $\chithree$.

\section{Temperature Dependence}

\begin{figure}
    \includegraphics[width=0.4\textwidth]{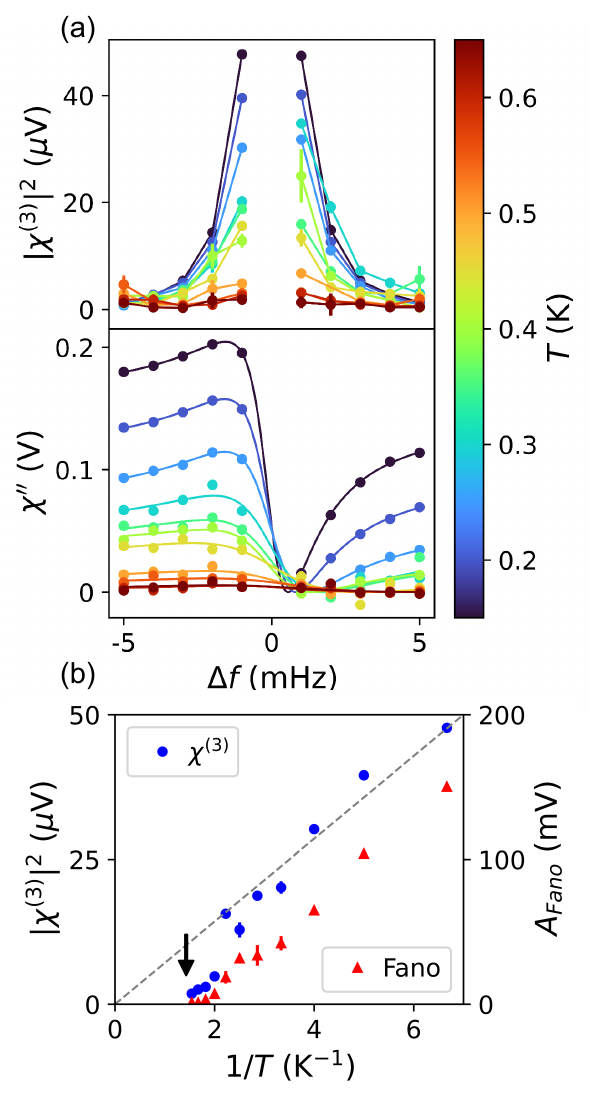}
    \caption{\label{fig:fig_3} Temperature dependence of the sample response. (a) The magnitude of the non-linear sample response (top) at $f = \fprobe + 2 \Delta f$ as a function of the detuning, $\Delta f$, for several $T$ (at $H_\perp$ = 0), and Fano resonance fits (bottom) of the imaginary part of the susceptibility, $\chi^{\prime\prime}$ at the probe frequency, $\fprobe$, as a function of probe detuning, $\Delta f$ for the same set of temperatures. (b) Amplitude of the non-linear signal $|\chithree|^2$ (blue circles), as defined by the magnitude of the $2 \Delta f$ signal at a detuning of $\Delta f = -1$ mHz plotted as a function of $1/T$ and the Amplitude of the corresponding Fano resonances (red triangles), with $1/T$ (dashed grey line) plotted for reference. Both the non-linear and Fano responses are almost completely suppressed for $T > 700$ mK.}
\end{figure}

The temperature dependence of the non-linear response reveals the pertinent spin cluster energy scales and dissipation mechanisms. We first extract the squared amplitude of the spectral components at $\fprobe + 2 \Delta f$, corresponding to $|\chithree|^2$, in Fig. \ref{fig:fig_3} (a) as a function of the probe detuning, $\Delta f \equiv \fprobe - \fpump$, for several temperatures. As expected, the signal dies rapidly as the probe frequency is detuned farther and farther away from the pump frequency. Similarly, the overall amplitude of the response decreases with increasing $T$. We compare the non-linear response with the imaginary part of the susceptibility at $\fprobe$ in Fig. \ref{fig:fig_3} (a). As in \cite{silevitch2019natcommun}, the signal at $\fprobe$ shows clear asymmetric Fano resonances, characteristic of quantum interference between multiple transition pathways. For a more detailed theoretical treatment of the nature of these Fano resonances we direct the reader to \cite{buchhold2020phys.rev.b}.

In Fig. \ref{fig:fig_3} (b), we plot the amplitudes of each signal as a function of inverse temperature, which show a rough $1/T$ dependence (dashed grey line) at low temperatures, with a high-temperature cutoff that almost completely suppresses both the non-linear response $\chithree$ and the amplitude of the Fano resonance for $T > 700$ mK. The coincidence of these two temperature cutoffs strongly suggests that the non-linear response is due, not to a bulk response of the entire dipole assembly, but to the response of the set of clusters that are both nearly resonant with the pump and probe fields and very weakly coupled to the rest of the spin bath. We contrast the approach here with measurements that probed the off-resonant non-linear susceptibility, which, due to the crystal symmetry, require a DC longitudinal field to bias the bulk response to exhibit any signal in the transverse polarization \cite{silevitch2007phys.rev.lett.}.

These observations resolve the paradox of the surprisingly weak field scales for the non-linearities. One would assume intuitively that the field scale needed to observe non-linear effects is of order the internal microscopic field scales of the sample. In \lihox, the nearest-neighbor longitudinal field is of order $\sim 1$ kOe, in stark contrast to the $< 1$ Oe pump fields used to drive our system into the non-linear response regime. The relevant energy scale to drive the system (or, more precisely, the clusters resonant with the pump/probe fields) non-linear, however, is the decoherence rate caused by dissipative phonon-scattering \cite{buchhold2020phys.rev.b} rather than the internal dipolar fields.

Dissipation, $\gamma$, due to phonon-scattering has been given a detailed treatment in the context of a toy model of a spin dimer/trimer in \cite{buchhold2020phys.rev.b}, and, in the case of weak heat-sinking (resulting in small phonon linewidths), scales linearly with T ($\gamma \sim T$). Furthermore, the predicted Fano resonance amplitude scales as $\frac{1}{\gamma}$, which predicts that the amplitude of the Fano resonance should scale as $\frac{1}{T}$. Our data scale roughly as $1/T$ at low temperatures, with a high-temperature cutoff at $\sim$ 700 mK, where the spin clusters effectively melt \cite{silevitch2019natcommun}. 

Previous susceptibility measurements have shown that the peak frequency in the imaginary susceptibility follows an Arrhenius law with an energy barrier of $E_B = 1.46$ K \cite{reich1990phys.rev.b}. This activation energy is of order the average dipolar interaction energy, and is likely what causes decoherence, and consequently what suppresses both the four-wave mixing and the quantum interference visible in the Fano resonances in the imaginary susceptibility at the probe frequency.

\section{Transverse Field Dependence}

A remarkable property of the \lihox $ $ system is that the quantum interference between excitation pathways can be tuned via the application of a DC transverse field. One measure of the tunable interference is the quantum phase between excitation pathways, which can be characterized by the Fano asymmetry parameter, $q$. The Fano lineshape exhibits a leftward-skewed resonance for $q>0$, and a rightward-skewed resonance for $q<0$, with a symmetric response at the point $q=0$. As has been demonstrated in previous work \cite{silevitch2019natcommun}, the $q=0$ point is associated with dissipationless cluster response ($\chi^{\prime\prime}(\fprobe) = 0$). 

\begin{figure}
    \includegraphics[width=0.4\textwidth]{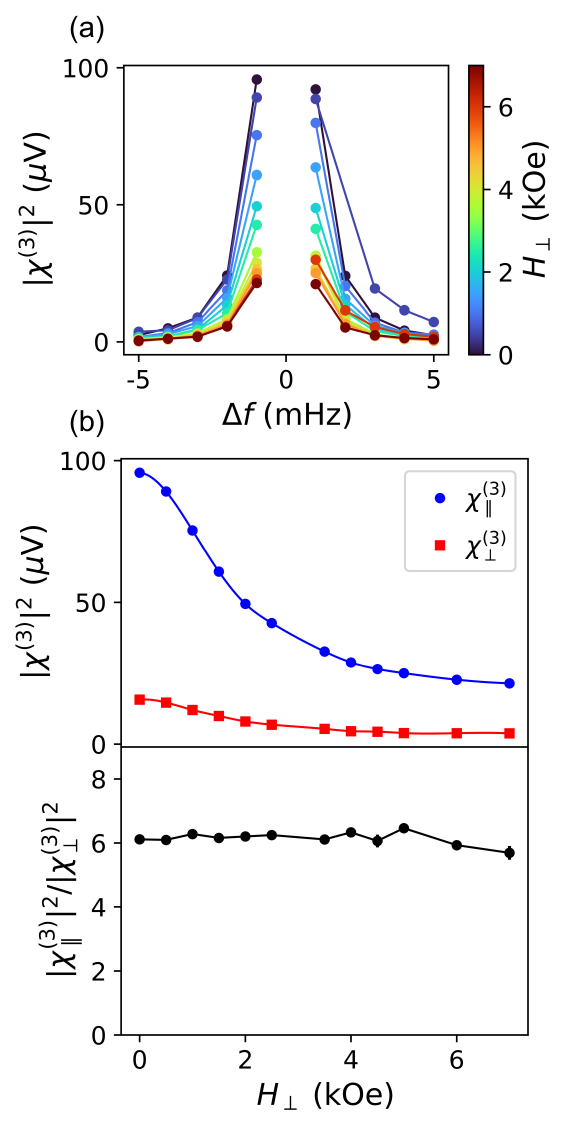}
    \caption{\label{fig:fig_4} Transverse field dependence of the sample. (a) The magnitude of the non-linear sample response at $f = \fprobe + 2 \Delta f$ as a function of the detuning, $\Delta f$, for several $H_\perp$ (at $T$ = 110 mK). (b) Amplitude of the non-linear response (top) as characterized by the signal at a detuning $\Delta f$ = -1 mHz as a function of $H_\perp$, for the longitudinal (blue circles) and transverse (red squares) sample polarization channels. Both channels show a monotonic decrease approaching a constant value at $H_\perp \geq$ 5 kOe, and the ratio of the non-linear responses (bottom) for the longitudinal and transverse channels. This ratio is approximately constant, and independent of $H_\perp$.}
\end{figure}

Similarly, the non-linear response $\chithree$ can be tuned via an external transverse field. We plot in Fig. \ref{fig:fig_4} the non-linear signal $|\chithree|^2$ as a function of the detuning between the pump and probe frequencies, for a set of external transverse fields $H_\perp$. Similar to the temperature dependence discussed above, the non-linear response exhibits a symmetric form, decreasing as the probe frequency is detuned away from the pump frequency. Furthermore, increasing the transverse field monotonically decreases the non-linear response.

To illustrate the $H_\perp$ dependence, we plot the non-linear response as characterized by the response at a fixed detuning $\Delta f = -1$ mHz in Fig. \ref{fig:fig_4} (b) for the longitudinal and transverse field channels. They both monotonically decrease with increasing $H_\perp$, approaching a constant value for transverse fields above a few kOe. As the strength of the transverse field is varied, it should theoretically shift both the cluster levels as well as the matrix elements $\langle \alpha | J^z | \beta \rangle$ connecting the cluster levels to one another via the AC longitudinal field. While it is impossible to do an \textit{a priori} calculation of the functional form of this transverse field dependence, as it would require knowing the exact spatial geometry of the clusters in resonance with the pump/probe fields, and then doing exact diagonalization on a cluster of 10s or even 100s of spins for various transverse fields, we see that the internal cluster rearrangements approach a point above a few kOe where the effects of the external transverse field are negligible. The non-linear signal approaching a constant value at high transverse fields suggests that that four-wave mixing is due to a combination of both coherent Raman scattering, which is suppressed as the cluster states are tuned out of resonance by the transverse field, and spontaneous Raman scattering, which is independent of the transverse field.

Comparing the evolution of the longitudinal and transverse polarization channels with $H_\perp$ provides additional information about the spin cluster dynamics. The two channels follow the same functional form. The ratio of the amplitudes of the longitudinal and transverse fields is a constant, independent of transverse field, as plotted in Fig. \ref{fig:fig_4} (b). We conclude that the transverse channel measures a smaller projection of the anisotropic cluster excitation into the plane transverse to the Ising axis. $H_\perp$ simply augments the matrix elements connecting the states, but not the observable magnetic moments of the dressed states themselves. While this may seem puzzling at first glance, it is known that quantum tunneling between Ising states occurs as a second-order perturbative effect through virtual transitions mediated by the higher-energy levels in the 17-dimensional crystal field manifold \cite{chakraborty2004phys.rev.b}. Hence, the observable magnetic moments are dominated by the zeroth-order term due to the magnetic moments of the Ising states. They are relatively insensitive to perturbations due to small amounts of mixing with the higher energy crystal field levels, while transitions between levels, which are dominated by the second-order perturbative term, are highly sensitive to small transverse fields.

\section{Conclusions}

We measure the non-linear susceptibility $\chithree$ of the disordered quantum Ising magnet, \lihofour, using a pump-probe technique, and a purpose-built vector susceptometer that permits access to the sample response both along and transverse to the Ising axis. We observe four-wave mixing due to coherent (anti-)Stokes Raman scattering processes. Based on an analysis of the temperature dependence of the non-linear response, we attribute the four-wave mixing to coherent (anti-)Stokes Raman scattering due to quantum interference between excitation pathways of spin clusters resonant with the pump/probe fields that are only very weakly coupled to the spin bath. The dependence of $\chithree$ on the external transverse field strength suggests that the four-wave mixing has contributions from both coherent Raman scattering, which is suppressed by transverse fields above a few kOe, and a spontaneous term that is independent of transverse field. Furthermore, the off-diagonal dipolar interactions between spins within a cluster break the crystal symmetry and allow for projections of the cluster excitations into the transverse plane, which are also independent of transverse field. These experiments, employing a Floquet drive to drive a quantum magnet system into a non-equilibrium state and using four-wave mixing techniques to probe the dressed states, marshal the tools of non-linear optics to discern the dynamics of magnetic spin clusters at $\sim$ 100 Hz frequencies.

\begin{acknowledgments}
We are grateful to G. Refael and P. Stamp for thought-provoking discussions. This work was supported by the US Air Force Office of Scientific Research, Grant \#FA9550-20-1-0263.
\end{acknowledgments}

\bibliography{Four_Wave_Mixing_Paper_Refs}

\end{document}